\documentclass[varenna]{cimento}
\usepackage{graphicx}
\usepackage{cite}
\usepackage{amsmath}
\numberwithin{equation}{section}
\title{Pattern formation and shocks in granular gases}
\subtitle{To appear in Proceedings of The
International School of Physics Enrico Fermi Course CLV "The Physics of 
Complex Systems (new advances and 
perspectives)" edited by
F. Mallamace and H. E. Stanley (IOS Press, Amsterdam, 2004).}

\author{Harry L.~Swinney,
E.C.~Rericha}
\institute{Center for Nonlinear Dynamics and Department of
Physics, University of Texas at Austin, Austin, Texas 78712 USA}

\begin{document}

\begin{abstract}
Granular media such as sand and sugar are ubiquitous in nature and
industry but are less well understood than fluids or solids.  We
consider the behavior of rapid granular flows where the transfer of
momenta by collisions dominates. The physics is quite different for
the opposite limit of static or slowly moving grains (e.g., sand
piles). To gain understanding of granular flows we consider two
problems that have been investigated with experiments, particle 
simulations and hydrodynamic theory:
vertically oscillating granular layers and flow past an obstacle.
Oscillating granular layers spontaneously form spatial patterns when
the container acceleration amplitude exceeds a critical value, about
2.5 times the gravitational acceleration.  Simulations with hard
spheres that conserve linear momentum and dissipate energy in
collisions are in qualitative accord with some but not all aspects of
the observed patterns.  It is necessary to include friction and
angular momentum conservation in the simulations to achieve
quantitative accord with observations.  

The applicability of a hydrodynamic theory to granular flows is not
obvious because for typical conditions the particle mean free path is
comparable to the length scale over which velocity and density fields
change; hence there is not the separation of scales needed to justify
a hydrodynamic approach.  However, we show that Navier-Stokes-like
equations describe well the density and temperature fields in
vertically oscillating layers, even though this system is far from
isothermal, incompressible fluid dynamics.

The second problem examined, flow past an obstacle, is also described
well by particle simulations for frictional dissipative
particles, but continuum simulations are only in qualitative
accord with the observations. We show that continuum theory fails
because it does not include friction between particles.  Finally, we
discuss how shock waves are common in granular flows since the the
speed of sound (pressure waves) in a granular gas is typically only a
few centimeters/second, while mean flow speeds are typically
meters/second.  Comparison of shocks in granular experiments and
simulations is made with shocks in ordinary gases.  Much more research
is needed to understand how shocks evolve in granular flows.

\end{abstract}
\maketitle

\section{Introduction}

Granular materials include sand, sugar, crushed coal, cereals,
pills, cosmetics, and asteroids. The transport, mixing, and
segregation of granular materials is important in the pharmaceutical,
mining, agricultural, metal, food, and energy industries. Even in the
chemical industry, the majority of the products are in granular rather
than liquid form~\cite{ennis94}. Thus a large engineering literature
has developed on ``powders and particulates''.  However, a basic
understanding of the physical mechanisms underlying the collective
behavior of particles in a granular medium is lacking. Granular media
can exhibit both solid and fluid properties (e.g., one can walk on a
beach or pour the sand from a bucket), but granular media are less
well understood than solids and fluids.  While fluids are processed
in industry with high efficiency, the efficiency of handling
(crushing, mixing, separation) of granular materials is estimated to
be well below optimum~\cite{ennis94}.

The scientific study of granular systems has a long history, including
a discussion by Galileo in his {\it Dialogues} and an 1831 study by
Faraday of convective motion of grains in heaps in vertically
oscillated granular layers~\cite{faraday}. In recent years there has
been a resurgence of interest in granular media among physicists,
thanks to Pierre-Giles de Gennes, who recommended in the early 1980s
to young French scientists that granular matter was an interesting
subject worthy of study~\cite{evesque89,fauve89,douady89,degennes99}. In the
past two decades there has been an explosion of interest in granular
systems.  A search in INSPEC on the word ``granular'' followed by
``system, medium, matter, flow, or gas'' yields 20 papers in the three
year period 1980-82, while a decade later the number of papers in a
three year period jumped to 112, and in 2000-2002 there were 691
papers on the subject.  The growth is not so dramatic for literature
searches using terms often used in industry (powder, particles,
particulates), but most of that work is empirical, and much of that
literature concerns pastes, soil, and fine particles (powders).  We
make no attempt to review the enormous literature on
granular/particulate matter but list a few reviews~\cite{campbell90, jaeger92,
jaeger96,rajchenbach00,goldhirsch03,goldshtein03,xu03}, 
books~\cite{bagnold54,bideau93,duran00a,poschel03a,
poschel03b}, and several journals: {\it Powder Technology}, {\it
Granular Matter}, {\it Particle Science and Technology}, and {\it 
Advanced Powder Technology}.

The key property distinguishing collisions of granular particles from
collisions of atoms in an ordinary gas is dissipation: collisions
between macroscopic grains are inelastic -- in the absence of external
forcing, the particles in a granular medium all come to rest. A
measure of this dissipation is the coefficient of restitution $e$,
which is the ratio of the relative normal velocity of two particles
after a collision to the relative normal velocity before
collision. Some representative values of $e$ for a relative normal
velocity of 0.1 m/s are 0.96 for hardened bronze, 0.85 
for aluminum,
and 0.3 for lead~\cite{goldsmith}; the values depend on the particle bulk and
surface properties in a complicated way (see, e.g., ~\cite{louge97}).

This chapter concerns rapid granular flows, which is called the
collisional regime to distinguish it from the quasi-static regime
where particles are at rest or nearly so. Much of the granular
literature concerns the quasi-static regime where inertia is not
significant and chains of particles in contact bear most of the
load. Understanding the development and evolution of these force
chains and the role of steric hindrance is often the focus of the
research on sand piles and other granular systems at rest or nearly at
rest. The collisional and quasi-static regimes are each difficult, but
the intermediate regime with some particles moving rapidly while
others are at rest is even more difficult. For an example, see the
contribution in this book on the formation of craters in a granular
medium~\cite{pica04,pica04a}.

We will consider situations where only contact forces are
important.  For small particles (less than$\approx$50 $\mu$m), 
electrostatic
and van der Waals forces become important. Further, air friction can be
significant, but for particles greater than about 1 mm in diameter, air
friction is often negligible if the velocity is not too large. 

An oft-studied example of a granular medium in the collisional regime
is a collection of particles in a vertically oscillating
container. Section 2 describes spatial patterns formed by such a
system. Since the number of particles in an experiment can be small (less
than one million), Newton's laws for the motion of these particles can
be directly implemented on a Personal Computer. Section 3 discusses
such Molecular Dynamics (MD) simulations for hard, spherical particles 
that are
characterized by a restitution coefficient and a frictional
coefficient.  Most theoretical analyses of granular flows examine
frictionless (smooth) inelastic spheres, but there exist no
frictionless macroscopic particles, just as there are no elastic
particles.  Molecular dynamics simulations show that realistic models
of rapid granular motions must include friction -- friction provides
another mode of dissipation and also results in reduced grain mobility
and a higher particle collision rate~\cite{moon04}. Molecular
dynamics simulations including friction describe experimental
observations on oscillating granular media very well, as we shall
describe in Section 3, but simulations without friction fail to
capture even qualitatively some important aspects of the observations.

Studies of oscillated granular layers have revealed localized
structures, ``oscillons", which are stable for a range of
container oscillation frequencies and amplitudes.  Section 4
describes the properties of oscillons and shows how they can be
considered as the basic building blocks of some extended spatial
patterns.

The continuum approach to granular flows is introduced in Section
5, where equations of motion derived by Jenkins and Richman for a
dilute dissipative fluid are presented. The Jenkins-Richman
equations are similar to the Navier-Stokes equations for a fluid
but are modified to include dissipative effects arising from
collisions between the grains. However, the equations do not
include the effect of friction in grain-grain collisions or
grain-container collisions.  As will be discussed, the application
of continuum equations to granular flows must be considered with
caution because, unlike an ordinary fluid, in granular flows there
is not a large separation between the microscopic and macroscopic
length and time scales.

In a granular fluid the speed of propagation of pressure
fluctuations (the sound speed) is only a few centimeters/sec, which is
very small compared to the 330 m/s sound speed in air.  (The
granular sound speed here refers to particles in vacuum; there is
no air and the only particles are the dissipative grains.)  Hence
the average streaming speed of granular flows generally exceeds the sound 
speed in the flow, and shock waves form whenever the
flow encounters an obstacle. These shock waves are the subject of
Section 6, where observations from experiments are compared with
molecular dynamics and continuum simulations.

We will conclude with a discussion in Section 7 that mentions some of
the many issues open in the understanding of systems of dissipative
particles.

\section{Patterns in vertically oscillated granular layers}

\begin{figure}
\begin{center}
\includegraphics[width=60mm]{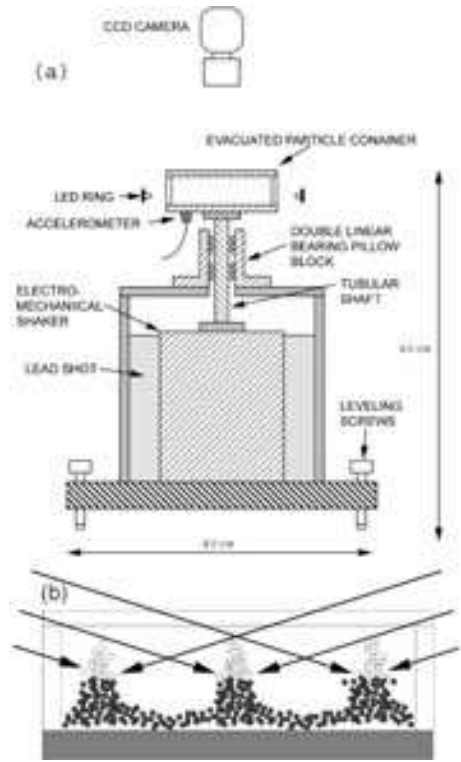}
\caption{(a) Electromechanical shaker system for studying pattern
formation in vertically oscillating layers. (b) Spatial patterns are
illuminated from the side by light incident at low angles.  When
viewed from above, high regions are bright and low regions are
dark. From \cite{umbanhowar00}.}
\label{apparatus}
\end{center}
\end{figure}

Consider a layer of non-cohesive grains in a container oscillating
sinusoidally with frequency $f$ and dimensionless acceleration
amplitude $\Gamma = 4\pi^2 f^2 A/g$, where $2A$ is the
peak-to-peak amplitude of the displacement of the container and
$g$ is the gravitational acceleration.  We consider patterns that
arise spontaneously, not from sidewall forcing or from
interstitial gas, but from correlations induced by multiple
collisions between the grains and between the grains and the
container bottom.  To minimize sidewall effects, the horizontal
dimensions of the container are made large compared to the layer
depth $h$.  The layer is illuminated from the side, as
fig.~\ref{apparatus} illustrates~\cite{umbanhowar00, bizon98}.

\begin{figure}
\begin{center}
\includegraphics[width=135mm]{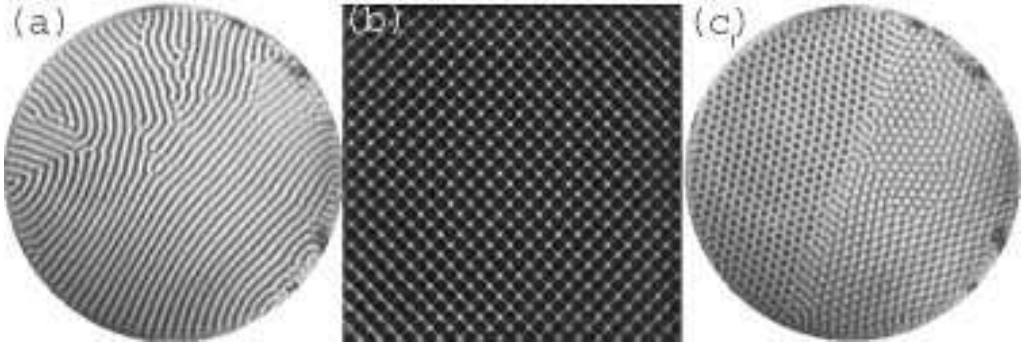}
\caption{Patterns in oscillating granular layers: (a) stripes, (b)
squares, and (c) hexagons (with different phases on the left and
right). The patterns oscillate at $f/2$, where
$f$ is the container oscillation frequency. Here $\Gamma$, $f$,
and $h/\sigma$ (ratio of the depth of the layer at rest to the diameter
of the particles) are given by (a) 3.3, 67 Hz, 7, (b) 2.9, 25 Hz,
4, and (c) and (d) 4.0, 67 Hz, 7. In (a) and (c) the diameter of
the container is 770$\sigma$ and in (b) the container is square, 
$1100\sigma
\times 1100\sigma$. The particles are bronze spheres 0.165 mm diameter.
(a) and (c) are from \cite{melo95} and (b) is from
\cite{goldman03}.} \label{patterns}
\end{center}
\end{figure}

For $\Gamma > 1$, on each cycle the layer loses contact with the
container, flies in the air, and then collides with the container.
However, the layer remains compact and flat until $\Gamma \approx
2.5$, where a standing wave pattern spontaneously forms.  A square
pattern forms at low frequencies and a stripe pattern at high
frequencies, as figs.~\ref{patterns}(a) and (b) respectively
illustrate~\cite{melo95, goldman03}.  The pattern is subharmonic, 
repeating every $2\tau$, where
$\tau=1/f$ is the container oscillation period; thus a ridge in a striped
pattern at an instant of time becomes a valley one container oscillation
period later.

A heuristic argument for the critical value of $\Gamma$ for the
onset of instability of a flat oscillating layer was given in
~\cite{bizon98}. The authors argued that a flat layer becomes
unstable when the collision occurs at the plate's lowest point.
Then a time $\tau/2$ is taken for the layer to free fall from its
highest point through a distance $2A$ to collide with the plate,
so that $2A=\frac{1}{2}g(\tau/2)^2=\frac{1}{8}g\tau^2$, which gives
$\Gamma_c=2.5$, in accord with the observed critical acceleration
for the onset of patterns.

The onset of patterns (squares or stripes) at $\Gamma \approx 2.5$ is
quite robust, independent of layer depth (depths
from about a monolayer to about $25 particle diameters \sigma$), container 
shape, and particle
properties (size, restitution coefficient, surface roughness,
material).  Most studies of patterns have been conducted for particles
of uniform size, but studies with a range of particle sizes have found
that the onset remains sharp for size distributions ranging up to
about 30\%~\cite{umbanhowar96}. Square and stripe patterns have also
been observed to form at $\Gamma \approx 2.5$ for irregular particles
such as rice grains and grass seed~\cite{umbanhowar96}.

\begin{figure}
\begin{center}
\includegraphics[width=100mm]{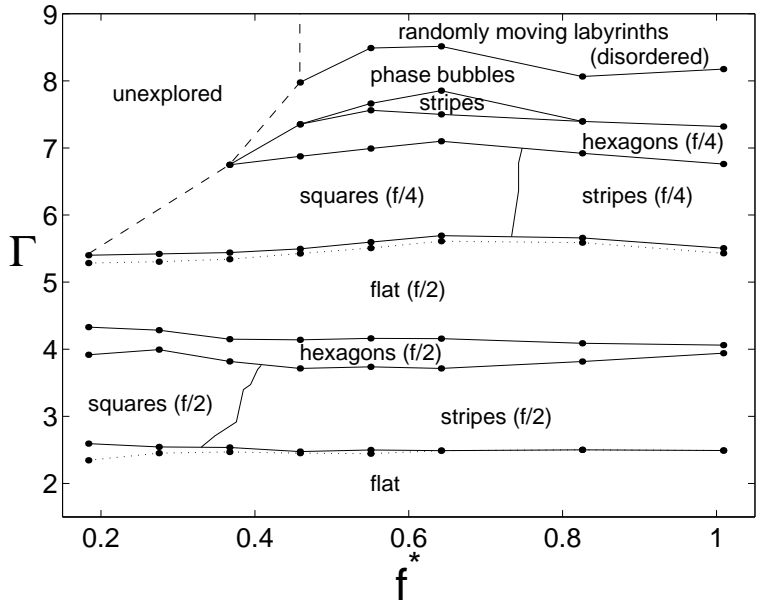}
\caption {Phase diagram for granular patterns observed in a
vertically oscillated container, as a function of the
dimensionless acceleration $\Gamma$ and dimensionless frequency
$f^* = f\sqrt{h/g}$. The transitions from a flat layer to squares
are hysteretic: solid lines denote the transition for increasing
$\Gamma$ while dotted lines denote decreasing $\Gamma$. (Bronze
spheres, $\sigma=0.165$ mm; layer depth, 5.0$\sigma$; container diameter,
770$\sigma$.) From~\cite{moon01}.} \label{phasediagram}
\end{center}
\end{figure}

The phase diagram (fig.~\ref{phasediagram}) shows the stability regions 
for different patterns as a function of $\Gamma$ and dimensionless frequency 
$f^*$, where $f^*=\sqrt{h/g}$ and $h$ is the depth
of the layer at rest~\cite{moon01}.  The transitions are well defined and 
are only weakly dependent on $f^*$. Except for the transition from a
flat layer to squares, the hysteresis is small or perhaps zero. We
know of no argument for the square to stripe transition as a
function of frequency with $\Gamma$ fixed, but we note that the
transition has been found to occur at $f^*\approx 
1/3$~\cite{umbanhowar00}.

\begin{figure}
\begin{center}
\includegraphics[width=60mm]{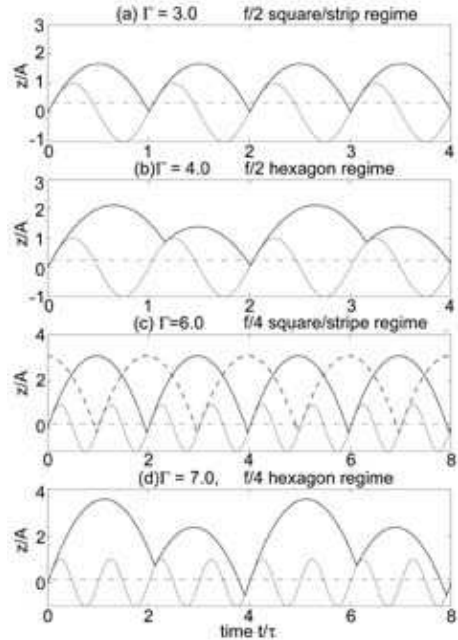}
\caption{Trajectory of a completely inelastic ball on an
oscillating plate. This is a model for the motion of the center of
mass of a granular layer.  The sinusoidal curve is the trajectory
of the plate.  The ball leaves the plate when the acceleration of
the plate becomes -$g$, that is, when the dot-dashed line
intersects the trajectory of the ball.  If the ball collides with
the plate above the dot-dashed line, as in (b) and (d), it leaves
the plate immediately. From \cite{moon01}.} \label{ball}
\end{center}
\end{figure}

Some insight into the dynamics can be gained from consideration of the
one-dimensional motion of a single completely inelastic ball on a
vertically oscillating plate~\cite{mehta90}. Such a model cannot of
course describe the 2D spatial patterns that form in granular layers
on an oscillating plate, but it does help in understanding the
transitions in behavior as a function of $\Gamma$~\cite{melo95,
moon01}. The inelastic ball motion, illustrated in fig.~\ref{ball},
models the center of mass motion of a granular layer at small
$\Gamma$, where the layer remains fairly compact and is highly
dissipative as a consequence of multiple collisions.  We will refer to
the inelastic ball as ``the layer", meaning the motion of the center
of mass of the granular layer.

For $\Gamma > 1$, the layer leaves the plate at the point in each
cycle when the plate acceleration exceeds $-g$. The layer continues in
free flight until it later collides with the plate. In the regime with
squares or stripes oscillating at $f/2$, the layer leaves and hits the
plate every cycle, as fig.~\ref{ball}(a) illustrates. At $\Gamma
\approx 4.0$, there is a bifurcation in the dynamics, illustrated by
fig.~\ref{ball}(b): successive trajectories now have different flight
durations, with a short trajectory initiated by a collision of the
layer with the plate whose acceleration greater than $g$, followed by
a long trajectory initiated by a collision of the layer with the plate
whose acceleration is less than $g$. This bifurcation corresponds to a
value of $\Gamma$ close to that for the onset of hexagonal spatial
patterns~\cite{melo95}, pictured in fig.~\ref{patterns}(c).  While
square or stripe patterns have the same appearance whether an image is
obtained at a time $t$ or $t+\tau$, hexagonal patterns at these two times
are different. The two phases are both present and are separated by a
phase discontinuity in fig.~\ref{patterns}(c) -- the pattern on the
right consists of a hexagonal array of dots, while the hexagonal
pattern on the left is cellular; the two patterns will be switched
one period, $\tau$, later.

Another bifurcation in the dynamics occurs for $\Gamma > 4.5$,
where the layer flight duration exceeds $\tau$, and the layer hits
the container every other cycle. Now the
velocity of the layer relative to the plate at the instant of
collision goes to zero (at about $\Gamma=4.6$), and the layer
makes a soft landing; not enough momentum is transferred from the
vertical to horizontal direction to form patterns.  When $\Gamma$
is increased above about 5.4, there is again a transition from a
flat layer to a pattern, stripes at low frequencies and squares at
high frequencies. However, now the pattern period is $f/4$ instead
of $f/2$.  Further increase in
$\Gamma$ leads to another bifurcation from trajectories with a
single period to successive trajectories with different flight
durations (fig.~\ref{ball}(d)). Again this bifurcation in the
dynamics of a ball corresponds to the bifurcation of the patterns
from squares or stripes to hexagons.

At larger $\Gamma$ the layer becomes sufficiently dilated so that
the inelastic single ball model no longer provides a useful
description of the dynamics. But even the $f/3$ and $f/6$ regimes
of the single ball model ($8 < \Gamma < 11$) have been
observed transiently in molecular dynamics simulations and
laboratory experiments~\cite{moon01}.

\section{Molecular Dynamics simulations}

Simulations of granular media using rigid particles, soft particles,
and Monte Carlo methods have been conducted by many researchers since
the 1980s~\cite{campbell85a,campbell90,bideau93}. We consider a granular 
layer
modeled as a collection of hard spheres that interact only through
instantaneous binary collisions. Between successive collisions the
particles move only under the influence of gravity.  This is known as
an Event Driven (ED) type of Molecular Dynamics (MD)
simulation~\cite{rapaport95}.  Linear and angular momentum are
conserved in collisions, while energy is dissipated through collisions
and surface friction. The interaction between particles is described
by the coefficient of restitution $e$, the coefficient of sliding
friction $\mu$, and the rotation coefficient of restitution
$\beta$. (The same values of $e$, $\mu$ and $\beta$ are used for
ball-wall collisions as for ball-ball collisions.) This collision
model was developed by Walton~\cite{walton93} and used in hard sphere
MD simulations conducted by Bizon~\cite{bizon98} and
Moon~\cite{moon01}, as we shall describe.

The restitution coefficient is often taken to be a constant in MD
simulations, but a constant value of $e$ can lead to ``inelastic
collapse", where particles undergo an infinite number of collisions in
a finite amount of time~\cite{mcnamara92,luding96,goldman98}. However,
physically $e$ must approach unity as the relative normal velocity
$v_n$ approaches zero (the elastic limit).  The simulations presented
here assume a form for $e$ that goes to unity in the $v_n \rightarrow
0$ limit: $e(v_n)=1-B{v_n}^\frac{3}{4}$ for $v_n < v_0$, where $v_0$
is a crossover velocity, and $e(v_n)=e_0=$ {\it constant} for $v_n >
v_0$. The approach of $e$ toward unity for low $v_n$ avoids inelastic
collapse, while the constant value $e_0$ at high $v_n$ is
computationally efficient. The MD results are insensitive to the form
assumed for $e(v_n)$ for $v_n < v_0$, as long as $e(v_n)$ increases to
unity as $v_n$ approaches zero~\cite{bizon98}.

In a collision the tangential impulse is given by $\mu$ times the
normal impulse, with a cutoff corresponding to the crossover from a
sliding contact to a rolling contact. The crossover ratio of the
relative surface velocity after collision to that before the collision
is given by $\beta$, which we fix at -0.35, as in ~\cite{walton93}.
The parameters $e_0$ and $\mu$ are set to 0.7 and 0.5, respectively,
values obtained by fitting MD simulation results to laboratory
observations for square patterns in layers of lead particles at one
set of conditions~\cite{bizon98}.

\begin{figure}
\begin{center}
\includegraphics[width=135mm]{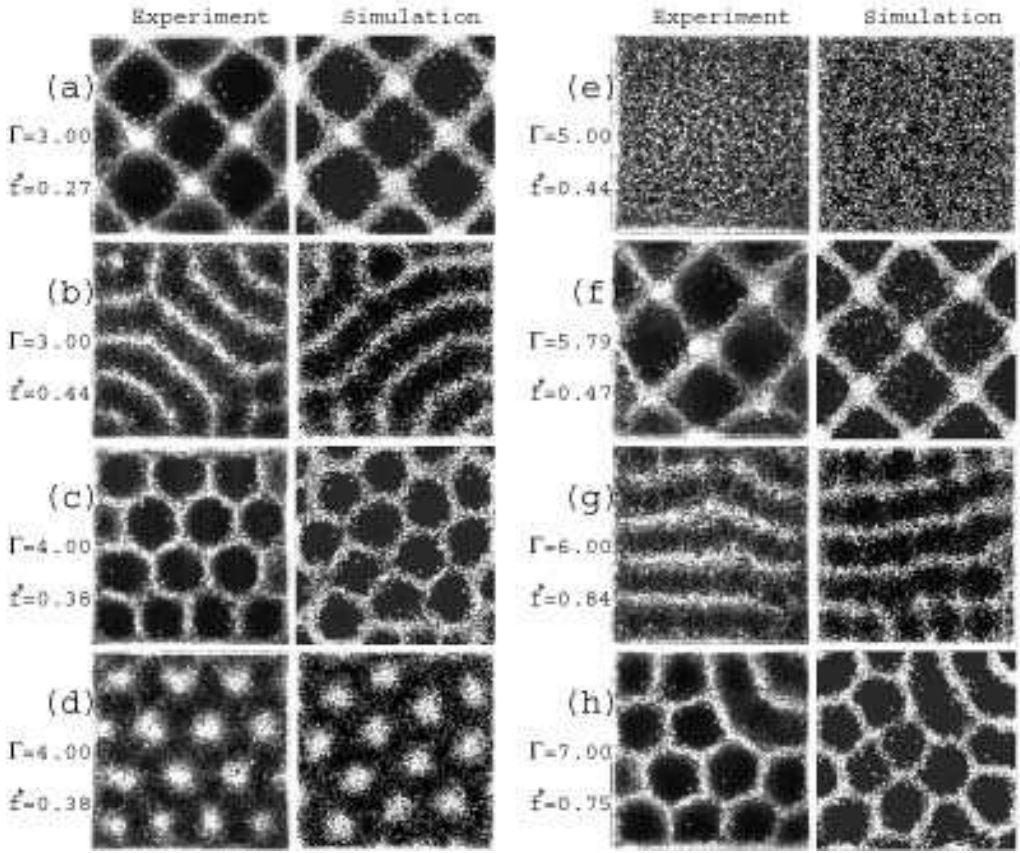}
\caption{Standing wave granular patterns in laboratory experiments
and molecular dynamics simulations for the same number of
particles, 60,000, which fill a $100\sigma \times 100\sigma$ container to 
a
depth of 5.4 layers: (a) squares, (b) stripes, (c) and (d)
alternating phases of hexagons, (e) flat layers, (f) squares, (g)
stripes, and (h) hexagons. Patterns (a)-(e) oscillate at $f/2$,
(f)-(h) at $f/4$.  The experiment used lead particles with
$\sigma=0.55$ mm. From~\cite{bizon98}.} \label{MDexpt} 
\end{center}
\end{figure}

Results from simulation and experiment are compared in
fig.~\ref{MDexpt} for seven values of ($f^*,\Gamma$).  At every point
in the phase diagram in fig.~\ref{phasediagram}, the results from the
MD simulation agree with experiment, not only in the form of the
pattern but also in the wavelength of the pattern.  The dispersion
relation relating the wavelength $\lambda$ to $f^*$ for layers of
varying depths collapse onto a single curve, $\lambda/h = 1 +
1.1{f^*}^{-1.32\pm0.03}$, when the container velocity exceeds a
critical value, $v_{gm}\approx 3\sqrt{\sigma g}$, where $v_{gm}$ 
corresponds
to a transition in the grain mobility ({\it gm}): for $v>v_{gm}$,
there is a hydrodynamic-like horizontal sloshing motion of the layer,
while for $v<v_{gm}$, the grains are essentially immoblile and the
stripe pattern apparently arises from a bending of the granular
layer~\cite{umbanhowar00}.

Most MD simulations and theoretical analyses of granular media
consider frictionless spherical 
particles~\cite{argentina02,nie02,khain03}. However, MD simulations
comparing the behavior of frictional and frictionless particles
indicate that the effect of friction cannot be mimicked by increasing
the dissipation (decreasing $e$)~\cite{moon04}; thus friction is not
merely an additional mechanism of dissipation. Even a small amount of
friction increases the overall dissipation significantly, not because
the frictional dissipation is significant in each collision, but
because the friction reduces the grain mobility and increases the
overall collision rate.  MD simulations with frictional particles
yield square and hexagonal patterns like those observed in experiments
(fig.~\ref{MDexpt}), while simulations without friction do not yield
square or hexagonal patterns, even if the restitution coefficient is
decreased to compensate for the absence of
friction~\cite{moon04}. Simulations of frictionless particles do
yield stripe patterns, but the critical $\Gamma$ for the onset of
patterns is smaller ($\Gamma_c \approx 1.9$) than for frictional
particles ($\Gamma_c \approx 2.5$), and the stripes formed by
frictionless particles are less robust than those formed by particles
with friction~\cite{moon04}.

\section{Localized structures and lattice dynamics}

\begin{figure}
\begin{center}
\includegraphics[width=135mm]{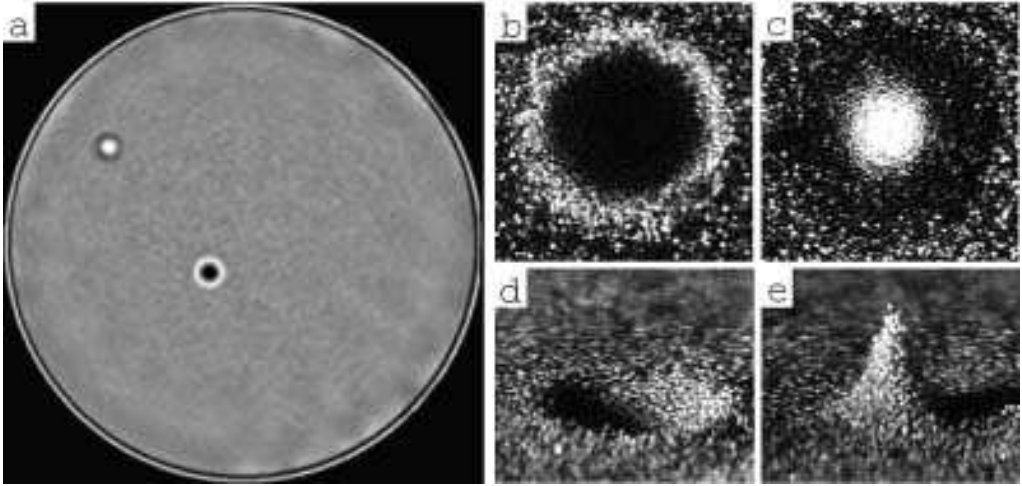}
\caption{(a) Snapshot of a container with two oscillons (viewed from 
above); one is
a peak (upper left) and the is other a crater (near the center). (b)
and (d), Close-ups of an oscillon crater viewed from the top and from
the side, respectively.  (c) and (e) Close-ups of an oscillon peak
viewed from the top and from the side, respectively.  Individual
bronze spheres ($\sigma=0.165$ mm) are discernible in (b)-(e).  ($f=25$ Hz,
$\Gamma=$2.45, layer depth $h=17\sigma$.)  From~\cite{umbanhowar96}.}
\label{oscillons}
\end{center}
\end{figure}

Localized stable standing wave structures dubbed ``oscillons" can occur
in oscillating granular layers when $\Gamma$ is decreased slightly
below the value corresponding to the onset of squares with increasing
$\Gamma$~\cite{umbanhowar96, umbanhowar98}. Top and side views of
oscillons are shown in fig.~\ref{oscillons}. An oscillon is a small,
circularly symmetric excitation that oscillates at $f/2$; during one
cycle of the container, it is a peak; on the next cycle it is a
crater. Unlike solitons, oscillons are stationary
(nonpropagating). Oscillons form with equal probability at all
locations in the container, and they live indefinitely. If the
container acceleration were increased slowly from rest to a value just
below the onset of squares, no oscillons would appear, but an oscillon
can be formed by a finite amplitude perturbation (a puff of air or a
poke with a rod).  When oscillons are obtained by decreasing $\Gamma$
from the regime with square patterns, the number of oscillons that
form is not fixed; as many as fifty oscillons were observed in the
experiment in fig.~\ref{oscillons}, but no oscillons occurred if $\Gamma$ 
was 
quasi-statically decreased~\cite{umbanhowar96a}.

\begin{figure}
\includegraphics[width=130mm]{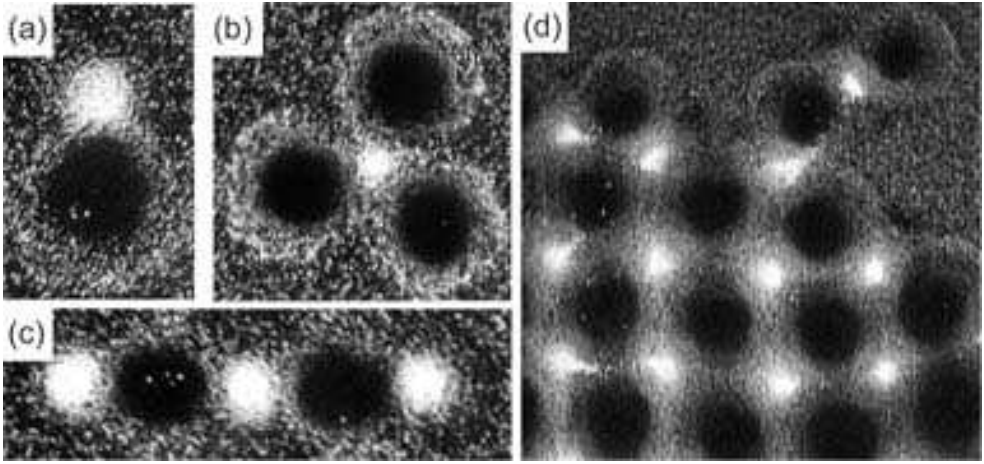}
\caption{(a) Dimer formed by two bound oscillons of opposite phase;
one period of oscillation of the container later the white peak will
have become a crater and the black will have become a peak. (b)
Tetramer formed of four oscillons. (c) Polymer chain of five
oscillons. (d) A square lattice grows by nucleating oscillons.  Individual
bronze spheres ($\sigma=0.165$ mm) are discernible in (a)-(c).  
From~\cite{umbanhowar96}}
\label{molecules}
\end{figure}

Oscillons of like phase show a repulsive interaction that has a range
not much larger than the diameter of an oscillon, while oscillons of
opposite phase that are closer than about 1.4 oscillon diameters
attract and form a stable dipole structure, as shown in
fig.~\ref{molecules}(a)~\cite{umbanhowar96,umbanhowar98}, and more
complex structures like the tetramer pictured in
fig.~\ref{molecules}(b) and the polymer chain in pictured in
fig.~\ref{molecules}(c).

Can oscillons be considered as building blocks (``atoms") of the
square lattice that forms with an increase in $\Gamma$? This view is
suggested by the observation of the formation of a square lattice as
$\Gamma$ is slowly increased (see fig.~\ref{molecules}(d)): an
oscillon seeds a square lattice by spawning oscillons, which adjust to 
form a
square array.  This observation suggests that the granular lattice
could be modeled as a system of coupled oscillon atoms, each of which
is comprised of hundreds of particles that are colliding hundreds
of times during each oscillation cycle.  Thus the lattice approach is
much simpler than the full description of all the particle motions and
simpler than a continuum fluid description of the granular medium.  As
we will now describe, the lattice picture is supported by an analysis
of the dynamics of the lattice.

\begin{figure}
\begin{center}
\includegraphics[width=100mm]{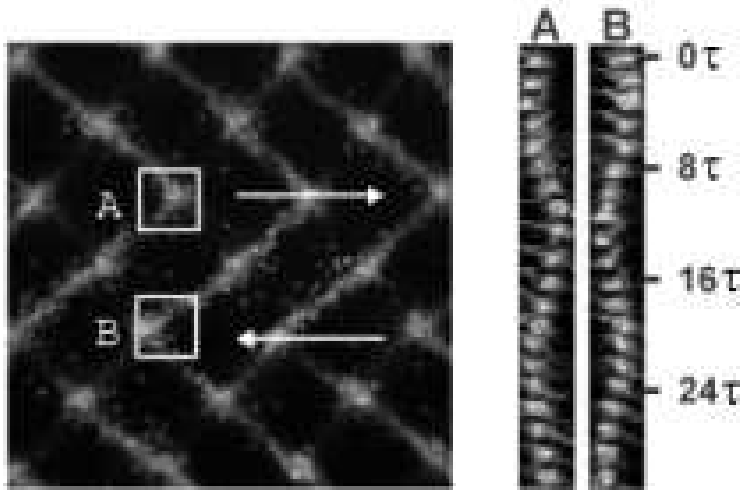}
\caption{Left: close up snapshot of a granular lattice
($\Gamma=2.9, f=25$ Hz, $h=4\sigma$). Right: time evolution of the
peaks in the boxes A and B in the left-hand image.  The peaks
oscillate out of phase with a frequency about twenty times smaller
than $f=1/\tau$. (from ~\cite{goldman03}).}\label{latticeosc}
\end{center}
\end{figure}

Close examination of the center of mass of a peak in a square lattice
reveals harmonic motion about the equilibrium position of the peak for
a wide range of $\Gamma$ and $f$~\cite{goldman03}; such a lattice
oscillation is illustrated in fig.~\ref{latticeosc}.  One test of the
conjecture that the lattice of peaks can be modeled by a lattice of
balls connected with springs is to compare the dispersion relation for
the two lattices. A time sequence of images of the granular pattern
was Fourier-transformed in space and then in time to obtain the
frequencies of oscillations of lattice modes with different wave
vectors. The frequency $f_L$ of the lattice modes as a function of $k$
(the magnitude of the wave vector) was found to be well described by
the dispersion relation for balls connected by springs, $f_L =
f_{BZ}|sin(ka/(2\sqrt{2})|$, where $f_{BZ}$ is the frequency at the
edge of the Brillouin zone and $a$ the lattice
spacing~\cite{goldman03}.  The lattice oscillation frequency $f_L$ is
typically an order of magnitude smaller than $f$ but depends on the
plate acceleration $\Gamma$ and frequency $f$.

\begin{figure}
\begin{center}
\includegraphics[width=135mm]{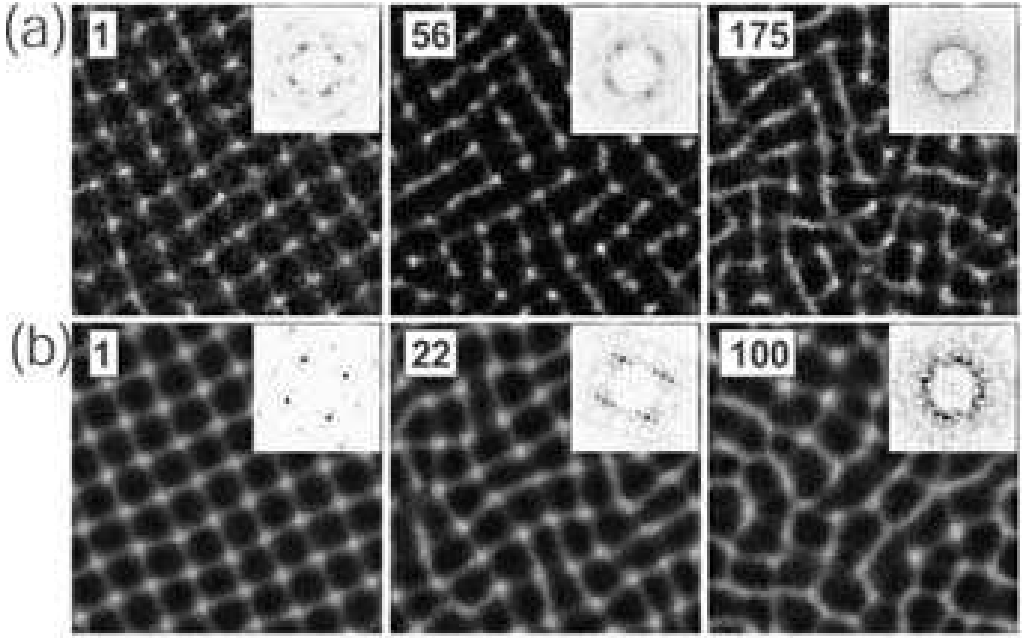}
\caption{Defect creation and melting of a square granular pattern
in a vertically oscillating granular layer: (a) experiment and (b)
molecular dynamics simulation. The insets show Fourier transforms
of the spatial patterns. In the experiment the plate oscillation
frequency ($f=32$ Hz) was modulated at the natural oscillation
frequency of the lattice (2 Hz), and at $t=\tau$ graphite powder was
added to the layer of bronze spheres. By $t=56\tau$ defects had
formed, and by $t=175\tau$ the lattice had melted. In the MD
simulation the friction coefficient was reduced from $\mu=0.5$ to
zero at $t=\tau$; by $t=22\tau$ defects had formed, and by $t=100\tau$ the
lattice had melted. ($\Gamma=2.9.$) From ~\cite{goldman03}.}
\label{melting}
\end{center}
\end{figure}

In a crystalline solid, defects form when the amplitude of
oscillation of atoms about their equilibrium position in the
lattice becomes large. To investigate the possible formation of
defects in a granular lattice, the plate frequency $f$ was
modulated at the lattice frequency $f_L$. For sufficiently large
modulation amplitude, defects formed, breaking the long range
order of the square lattice.  It was found that the amplitude of
the lattice oscillations could be increased further by adding a
lubricant (graphite powder) to reduce the friction between
particles. The result was that the granular lattice
melted~\cite{goldman03}: the spatial Fourier transform became a
circular ring (about $k=0$) rather than sharp peaks, as
fig.~\ref{melting} illustrates.

Lattice oscillations and defect formation have also been studied in MD
simulations, where friction can be easily varied or set to zero.
First a square lattice was simulated with $\mu=0.5$, as described
previously and illustrated in the first panel of
fig.~\ref{melting}(b). Then $\mu$ was set to zero, and defects were
observed to form quickly as the lattice oscillation amplitude
increased.  Finally the lattice melted, as illustrated by the last
panel of fig.~\ref{melting}(b).

Studies of melting in two-dimensional solids have shown that
melting occurs when the Lindemann ratio,
$\gamma=\langle|u_m-u_n|^2\rangle/a^2$, exceeds
0.1~\cite{bedanov85,zheng98}. Here $u$ corresponds to the
displacements of atoms from equilibrium lattice sites, $a$ is the
lattice constant, and the average is taken over all nearest
neighbors $m$ and $n$.  The Lindemann ratio was computed for the
granular lattice in the MD simulation, and it was found that when
the friction coefficient $\mu$ was decreased, $\gamma$ increased.
Further, when $\gamma$ reached the value 0.1 (which happened for
$\mu=0.1$ for the conditions of the simulation), the granular
lattice melted, in accord with the result for crystalline
solids~\cite{goldman03}.

\section{Continuum Description}

Section 3 introduced Molecular Dynamics simulations as a useful tool
in describing granular flows.  This technique models the system on a
microscopic level, evolving individual particle trajectories using
Newton's laws and computing the effects of each collision.  Averaging
over many collisions and particle trajectories gives the macroscopic
behavior of the flow.  A complementary method for understanding
granular flows is to model the macroscopic motion directly by a
continuum field theory that describes the bulk motion of the flow in
terms of the density, velocity and temperature fields.  Unlike MD
simulations, the continuum approach is not limited by particle number.
A personal computer currently contains enough memory for useful MD
simulations of laboratory experiments.  However, industrial processes
contain billions of particles, far outside the abilities of MD
simulations. Another reason that a continuum approach is attractive is
that it could exploit tools such as stability analysis, amplitude
equations, and perturbation theory, which have been developed through more
than a century of research on the the Navier-Stokes equations and
other partial differential equations.

Granular flows present many difficulties in developing a continuum
theory~\cite{kadanoff,tan98}.  Continuum theory requires a separation
of length and time scales: variations over space should be small and
occur over long distances, so that the behavior of local collections
of individual particles can be averaged and replaced with small fluid
elements.  Changes in time for the flow should occur for times long
compared to the mean time between particle collision so that particles
moving between fluid elements do not affect the average values in a
fluid element.  Unfortunately, inelastic collisions between particles
create an inherent lack of scale separation~\cite{tan98,
goldhirsch03}.  Sufficient separation of scales may only be present
for granular flows in the specific circumstances of low density
and low dissipation~\cite{kadanoff99,tan98,goldhirsch03}.

The derivation of the continuum equations from kinetic
theory makes assumptions about the underlying statistics of granular
flows, assumptions which have not been verified by MD simulations.  
For instance, the velocity distribution function is assumed to have a
steady state functional form that is nearly Gaussian.  Since granular
flows are dissipative, a steady state distribution function can only
be achieved in the presence of forcing. Granular experiments have
yielded velocity distributions that depend on the forcing
characteristics and experimental
geometry~\cite{olafsen99,kudrolli00,veje99,rouyer2000}.  Also, most
derivations of continuum equations assume Boltzmann's molecular chaos 
(particle velocities before collisions are uncorrelated), but strong 
velocity correlations have been found in MD simulations
~\cite{moon01b}.

\begin{figure}
\begin{center}
\includegraphics[width=135mm]{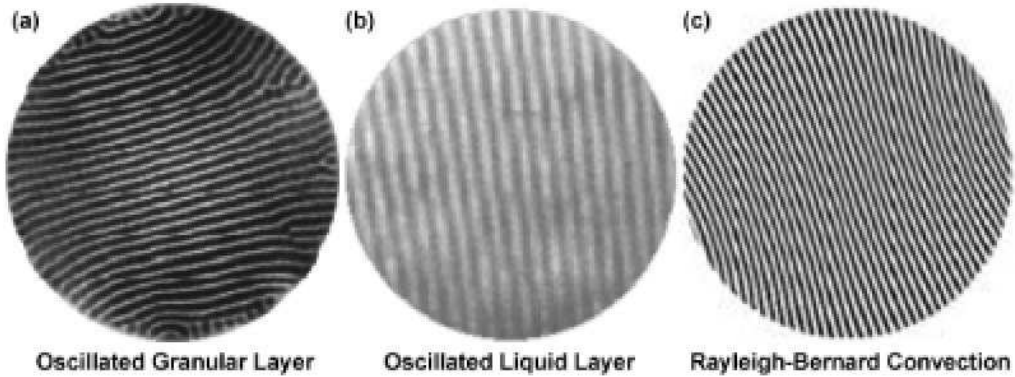}
\caption {Forced granular materials produce qualitatively similar   
patterns as forced fluids: (a) stripe pattern formed by a vertically
oscillated granular layer~\cite{melo95}, (b) stripe pattern formed by
a vertically oscillated layer of water~\cite{kudrolli96b}, (c) stripe
pattern formed in thermal convection of a fluid ($CO_2$)~\cite{plapp}.}
\label{faraday_waves}
\end{center}
\end{figure}

Despite the reservations regarding a continuum approach in granular   
media, observations of granular media have revealed many phenomena   
similar to those observed in continuum systems. For example, the
stripe patterns shown in fig.~\ref{faraday_waves}(a) look like
those in vertically oscillated liquid layers~\cite{kudrolli96b}
(fig.~\ref{faraday_waves}(b)), chemical reaction-diffusion
systems~\cite{quyang91}, Rayleigh-B\'enard convection in
fluids~\cite{behringer85}(fig.~\ref{faraday_waves}(c), and liquid 
crystals~\cite{dubois78}.  

\begin{figure}
\begin{center}
\includegraphics[width=120mm]{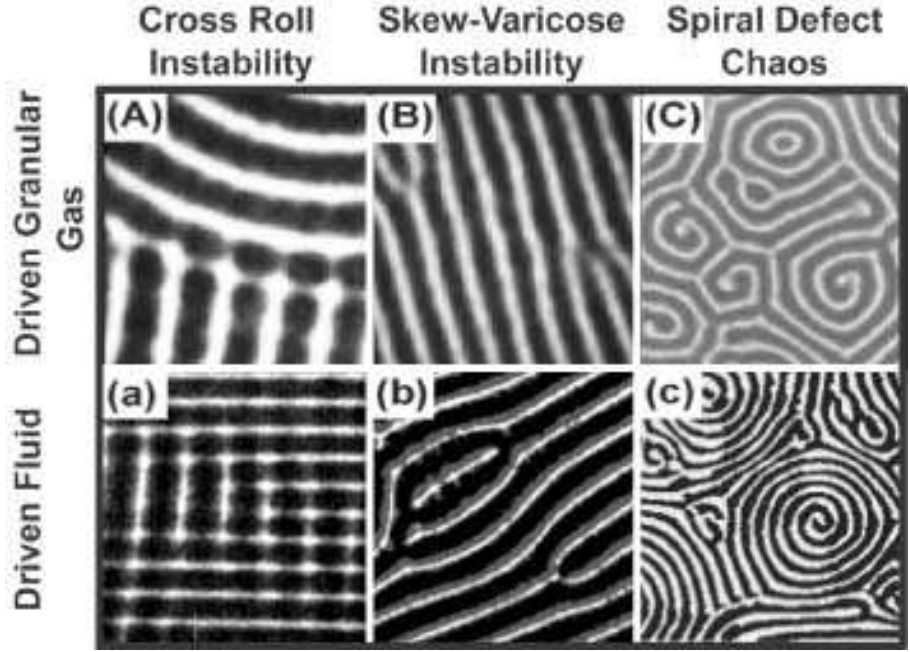}
\caption {Instabilities of patterns found in oscillating granular
layers and Rayleigh-B\'enard convection in a fluid.  Cross roll
instability in stripes: (A) Vibrated granular layer~\cite{debruyn98}  
and (a) Rayleigh-B\'enard convection~\cite{busse71}.  Skew varicose  
instability in stripes: (B) Granular layer~\cite{debruyn98} and (b)   
Rayleigh-B\'enard convection~\cite{assenheimer93}.  Spiral defect 
chaos in: (C) vibrated granular layer~\cite{debruyn02} and (c)
Rayleigh-B\'enard convection~\cite{assenheimer93}.}
\label{balloon}
\end{center}
\end{figure}

Not only are the patterns similar for granular and continuum systems,
but also some the same pattern instabilities have been observed.  For
example, when the wavenumber of parallel convection rolls (stripes) in
a Rayleigh-B\'enard convection becomes small, an instability leads to
the formation of cross rolls with a larger wavenumber that are
perpendicular to the original rolls~\cite{debruyn98,cross93}; the same
instability has been observed for stripes in oscillated granular
layers, as fig.~\ref{balloon} illustrates. The cross rolls invade the
region of small wavenumber stripes such that, after sufficient time,  
the region contains a pattern of straight stripes perpendicular to the
original pattern and with a larger wave number.

Granular stripe patterns also exhibit a skew varicose instability like
that in convection roll patterns (fig.~\ref{balloon}). When the local
wavenumber becomes too large, an initially straight pattern of stripes
will develop a distortion which evolves into a dislocation defect.
The defect propagates away; destroying one of the stripes and
decreasing the local wave number of the pattern.  The stability of the
stripe pattern in fluid convection is well described by amplitude
equations derived from the Navier-Stokes equations for
fluids~\cite{cross93}.  That the granular pattern shows the same
behavior strongly suggests a continuum description for the vibrated
system is applicable.

Aspects of the phase diagram 
for granular patterns 
(fig.~\ref{phasediagram}) have been reproduced by amplitude equation 
models.  For example, a phenomenological continuum model requiring that 
the mass of the 
layer is conserved locally, produces stripe, square, and 
oscillon-like patterns similar to those found in 
experiment~\cite{tsimring97, cerda98}.  A continuum, shallow water like 
model of the 
granular layer captures the patterns and yields a dispersion relation 
which 
agrees with experiment~\cite{eggers98}.   The success of these 
and other models~\cite{shinbrot97,sakaguchi97,rothman98,venkataramani98} 
provides further motivation 
for considering continuum equations derived for a granular gas.    

Additional evidence for the applicability of continuum theory to
granular media is provided by a recent study of noise in vertically
oscillating granular layers. In the Rayleigh-B\'enard system below the
onset of convection, thermal noise has been found to drive noisy
transient disordered waves with a characteristic length scale.  The
intensity and coherence of these modes increases as the transition
from conduction to convection is approached~\cite{oh03}.  This 
behavior is well-described by the fluctuating hydrodynamic theory of  
Swift and Hohenberg~\cite{swift77}.  Remarkably, the same noisy
incoherent modes are observed just below the transition from a flat   
vertically oscillating granular layer to a square
pattern~\cite{goldman04}. The Swift-Hohenberg continuum theory
describes the observations for the granular system very well, even
though the noise is not thermal noise, which is many orders of
magnitude too small; apparently the noise arises from the fluctuations
due to the small number of particles ~\cite{goldman04}.

Fired by the promise of quantitative predictive power and encouraged
by the qualitative similarity of granular flows to fluid flows,
researchers have proposed various continuum descriptions for rapid
granular
flows~\cite{haff83,ahmadi83,lun87,jenkins85,goldshtein95,sela96,brey97a}.
This section focuses on one such description~\cite{jenkins85} and
compares results from it to MD simulations and a granular flow experiment.

Jenkins and Richmann derived a set of inelastic continuum equations in
a manner similar to the derivation of the Navier-Stokes
equations~\cite{jenkins85}.  This approach begins with the single particle
distribution function $f^{(1)}\left({\bf r},t\right)$, which gives the
probability of finding a particle at a position ${\bf r}$ with a  
velocity ${\bf v}$ at a given time $t$.  Integrating $f^{(1)}$ over
all possible velocities gives the local number density, $n({\bf r},t)$. 
The 
ensemble averaged value of any particle property $\psi$ is determined
by

\begin{equation} <\psi>={1\over n}{\int \psi({\bf v}) f^{(1)}({\bf v},{\bf
r},t)d{\bf v}}. \end{equation}

The Boltzmann equation describes how $f^{(1)}$ changes in time.
Particles can move in and out of volume elements due to streaming
motion; particle velocities can change in response to external forces
${\bf F}$; or particles can be scattered out of elements by
collisions.  The time rate of change for $f^{(1)}$ is given by

\begin{equation} {\partial f^{(1)}\over\partial t}+{\bf
v}\cdot\nabla_{\bf r} f^{(1)}+{\bf F}\cdot\nabla_{\bf
v}f^{(1)}=\Theta(f^{(1}),\label{eq:boltz}\end{equation}

\noindent where $\Theta(f^{(1)})$ is the collision operator. Collisions 
are
considered to be binary, frictionless, and inelastic with a constant 
coefficient of restitution $e_0$~\cite{jenkins85}.  Integrating eq.
~\eqref{eq:boltz} yields the balance law for the number density,

\begin{equation} {\partial n\over\partial t} +
\nabla\cdot(n\mathbf{u})=0, \end{equation}

\noindent where ${\bf u}({\bf r},t)=(1/n){\int {\bf v}f^{(1)}({\bf r},{\bf 
v},t)d{\bf v}}$ is
the local average velocity.
Multiplying by the velocity and then integrating gives the balance law
for momentum,

\begin{equation}n\left( {\partial\mathbf{u}\over \partial
t}+\mathbf{u}\cdot\nabla\mathbf{u} \right) =
\nabla\cdot \underline{\mathbf{P}}- n
g{\mathbf{\hat{z}}} , \label{eq:momentum}\end{equation}

Finally, multiplying by ${\bf v^2}$ and integrating gives the balance
law for the energy, where the {\it granular temperature} $T$ is 
proportional to the average kinetic
energy of the random motion of particles,

\begin{equation} T=1/3\left(<{\bf v^2}>-<{\bf v}>^2\right), \end{equation}

\begin{equation}{3\over2}n\left({\partial T\over \partial t}+    
\mathbf{u}\cdot\nabla T\right) =
-\nabla\cdot \bf{q}\rm+\underline{\bf{P}}:\underline{\bf{E}}-{\gamma}.
\label{eq:temperature}\end{equation}

\noindent The granular temperature $T$ is many orders of magnitude greater 
than the Boltzmann temperature: thermal fluctuations are negligible 
($mg\sigma>>k_BT_B$).

A series of approximations is required in order to derive the form of
the pressure tensor ${\underline{\bf P}}$, the velocity gradient
tensor $\underline{\bf{E}}$, and the heat flux ${\bf q}$.  One assumes
that $f^{(1)}$ is nearly Gaussian, that spatial derivatives of $n$,
${\bf u}$, and $T$ are small, and that $(1-e_0)$ is small.  With these
assumptions, the components of the velocity gradient tensor
$\underline{\bf{E}}$ are given by: $E_{ij}={1\over2}\left({\partial_j
u_i}+{\partial_i u_j}\right)$.  The components of the stress tensor 
$\bf\underline{P}\rm$ are given by the constitutive relation:
\begin{equation} P_{ij}=\left[ -p + (\lambda
-{2\over3}\mu)E_{kk}\right] \delta_{ij}+2\mu
E_{ij}, \end{equation}
and the heat flux is given by Fourier's law:
\begin{equation}\mathbf{q}=-\kappa \nabla T.\end{equation}

The transport coefficients are fully determined and are the same as for a
dense gas of hard spheres.  The bulk viscosity is given by
\begin{equation}\lambda={8\over 3\sqrt{\pi}}n\sigma T^{1/2}
G(\nu),\end{equation}

\noindent the shear viscosity by
\begin{equation}\mu={\sqrt{\pi}\over6}n\sigma
T^{1/2}\left[{5\over16}{1\over G(\nu)} + 1 +
{4\over5}\left(1+{12\over\pi}\right)G(\nu)\right], \end{equation}

\noindent and the thermal conductivity by
\begin{equation}\kappa={15\sqrt{\pi}\over16}n\sigma
T^{1/2}\left[{5\over24}{1\over G(\nu)} + 1 +
{6\over5}\left(1+{32\over9\pi}\right)G(\nu)\right], \end{equation}

\noindent where
\begin{equation} G(\nu)=\nu g_0(\nu),\end{equation}
and the radial distribution function at contact, $g_0$,
is~\cite{goldshtein96}:

\begin{equation}g_0({\nu})=\left[ 1- \left(
{\nu\over\nu_{max}}\right) ^ {{4\over3}\nu_{max}}
\right]^{-1},\end{equation}
where $\nu$  is the volume fraction of the flow and $\nu_{max}=0.65$ is
the 3-dimensional random close-packed volume fraction.

The only difference between these equations and those for an elastic gas 
is $\gamma$ in eq.~\eqref{eq:temperature}, which accounts for
the temperature loss due to inelastic collisions:

\begin{equation} \gamma = {12\over \sqrt{\pi}} (1-e_0^2) {n
T^{3/2}\over\sigma} G(\nu). \end{equation}

The system is closed by an equation of state, proposed by Goldshtein
{\it et al.} in~\cite{goldshtein96},

\begin{equation} p=n T \left[ 1+2(1+{e_0}) G(\nu)\right].
\label{eq:state}\end{equation}

Direct experimental verification of the inelastic continuum equations
has been slow in coming due to the complexity of solving
the equations and also due to difficulties in finding an appropriate
experimental system ~\cite{campbell90}.  The presence of strong
gradients in granular materials~\cite{tan98,goldhirsch03} adds additional 
difficulty to solving continuum equations.  For instance, simulations of 
the vertically vibrated layer find that the temperature varies by three 
orders of magnitude throughout the cycle~\cite{bougie02}.  Thus, unlike 
most Navier-Stokes
simulations, the transport coefficients ($\lambda$, $\mu$ and
$\kappa$) cannot be treated as constants, but must be recomputed at
each grid point at every time step.  Additionally, a complete set of boundary
conditions for granular flows is still not established and this remains an 
active
area of research~\cite{silbert02,soto03,jenkins92,jenkins99,jenkins97,
brey00}.  Without the correct boundary conditions, numerical solutions
can be unstable and are not guaranteed to converge to a correct solution 
in the bulk.
For a good discussion on the difficulties in determining the correct
boundary conditions, see Goldhirsch's review
paper~\cite{goldhirsch03}.

In a 1990 review paper~\cite{campbell90}, Campbell made a resounding  
call for granular flow experiments to make quantitative tests of the
inelastic continuum approach.  The application of new technologies   
such as particle tracking in two and three dimensions is now making these
measurements feasible, but granular flow experiments still present
technical challenges.  Plugs develop in pipe
flow~\cite{aider99,poschel93}, wall effects dominate in 
quasi-two-dimensional experiments~\cite{rericha02}, and detailed bulk flow   
measurements are difficult to make in fully three dimensional  
experiments ~\cite{bocquet01}.

The distinguishing feature of granular flows is that inelastic
collisions dissipate energy.  Without an external source of energy,  
the granular temperature decays to zero with all particles coming to rest. 
Experimentally, energy can only be put into the flow through the
boundaries.  Shock waves serve as a mechanism to deliver energy from
the boundary to the bulk of the flow.  Studying the balance between
the transfer of energy by shock waves and the energy dissipation
through inelastic collisions is important in understanding granular 
flows~\cite{goldshtein03}.  In the next section we present two studies
of shock waves as a test for the inelastic continuum equations.

\section{Shock Waves in Granular Materials}
The sound speed (the speed a pressure wave travels) in a granular
medium is typically much smaller than the streaming velocity. Hence
shocks are common in granular media. For example, imagine pouring sand
out of a bucket.  Gravity accelerates the flow downward, creating an
average velocity $U$ that reaches 100 cm/s after the sand has fallen by 
only 5 cm.  In contrast, the sound speed $c$ in the granular gas
becomes small, typically 10 cm/s, as multiple particle collisions cool
the gas, reducing the random velocities of the particles.  The
simple act of turning over a bucket full of sand can easily generate a
supersonic flow with Mach number=$c\over U$=10.  (A flow with Mach
number greater than unity is supersonic.)

The sound speed in a granular gas can be determined from thermodynamic 
relations,

\begin{equation}c=\sqrt{\left({\partial
P\over\partial\rho}\right)_S}=\sqrt{{c_p\over c_v}\left({\partial
P\over\partial\rho}\right)_T},\end{equation}
where $c$ is the speed of a sound, $\rho$ is the local density, 
$S$ is
the entropy, $c_p$ is the specific heat at constant pressure, and $c_v$ is
the specific heat at constant volume.  For a dense inelastic gas, $c$
is given by~\cite{savage88}:

\begin{equation}c=\sqrt{T\chi\left(
1+{2\over3}\chi+{\nu\over\chi}{\partial\chi\over\partial\nu}\right)
},\end{equation}\label{eq:speed}
\noindent where $\chi = 1+2(1+e)G(\nu).$

A shock forms when a supersonic flow encounters an obstacle. The shock
separates two regions of the flow, an undisturbed region that is
unaware of the obstacle, and a compressed region which has adjusted to
fit the boundary conditions at the obstacle.  The compressed region
has a higher temperature and smaller velocity than the undisturbed
region.  In an ideal fluid with no viscosity, heat conduction, or
dissipation, a shock is a zero-width surface of discontinuity.  In a
non-ideal fluid the shock has a finite width on the order of a
particle's mean free path in the fluid~\cite{andersonbook}.

When a fluid with velocity $U>c$ impinges perpendicularly onto an
obstacle, a {\it normal} shock forms and propagates in the $-U$
direction.  Section 6.2 will discuss the formation and propagation of
normal shocks in a vertically vibrated granular layer where
the flow fields and thus the shocks are
highly time dependent~\cite{bougie02}. If, instead, the fluid
velocity and the obstacle are not perpendicular, an {\it oblique}
shock forms and propagates into the flow at an angle and with a speed 
determined by the local flow values.  Section 6.1 describes an  
oblique shock formed in a steady state laboratory flow past a
wedge~\cite{rericha02}.

\subsection{Steady state flow past an obstacle}

We now describe an experimental study of shocks in a time-independent
flow, and then the experimental results will be compared to two
simulations: an event-driven molecular dynamics simulation, similar to
those discussed in Section 3, and a two-dimensional, finite-element
solution of the inelastic continuum equations presented in Section 5.

In the experiment, stainless steel spheres (particle diameter
$\sigma=1.2$ mm) fell under gravity past a wedge sandwiched between two
glass plates separated by $1.6\sigma$.  The particles were initially
distributed uniformly on a conveyor belt.  As the conveyor turned,
particles fell off into a hopper that guided the particles into the   
cell formed of the closely spaced plates; the wedge was located a
distance of 42$\sigma$ below the top of the cell.  The positions and
velocities of the particles were determined from high speed images of
the falling particles, and data from many thousands of particles were
averaged to obtain the time-independent velocity, volume fraction, and
temperature fields.  The average free stream speed of sound determined
from the measurements for flow incident on the wedge was 0.09 m/s.   
The flow entered the top of the cell with a Mach number of 7 and accelerated
under gravity to a Mach number of 12 at the tip of the wedge.

\begin{figure}
\begin{center}
\includegraphics[width=135mm]{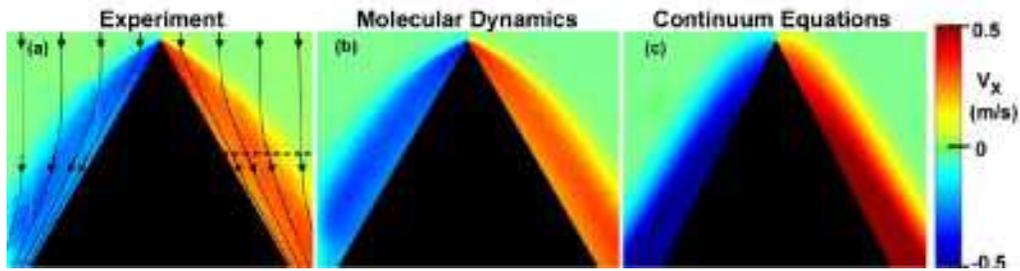}
\caption {Horizontal component of the velocity field of a granular
flow incident downward on a wedge, determined by three methods: (a) 
experiment, (b) MD simulations, and (c) integration of inelastic  
continuum equations.  Each picture shows a region 130$\sigma$ by      
104$\sigma$.  The solid lines with arrows denote streamlines.
Quantitative comparisons along the dashed line in (a) are shown in  
figs.~\ref{wp_friction} and~\ref{wp_no_friction}. 
(From~\cite{rericha02}).}\label{wedge_shock}
\end{center} \end{figure}

The horizontal velocity field measured in the experiment is shown in 
fig.~\ref{wedge_shock}(a).  A shock separates the undisturbed region,
where the horizontal velocity is nearly zero, from the compressed
region, whose stream lines follow the flow around the obstacle.
Because of gravity and inelasticity, the shock does not extend out at
a constant angle but curves towards the wedge.
\begin{figure}
\begin{center}
\includegraphics[width=60mm]{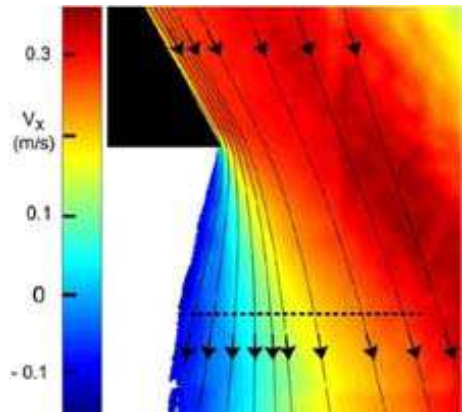}
\caption{The horizontal velocity field measured for the expansion fan
that formed when the supersonic granular flow reached the bottom of the
wedge.  The solid lines indicate selected streamlines.  The total
height of the region shown is 55$\sigma$.  The white region below the
wedge had too few particles for the velocity to be determined. (From
~\cite{rericha02}) }\label{expansion}  
\end{center} 
\end{figure}

At the bottom of the wedge the compressed gas expands in a fan-like  
structure as the volume available to the flow increases
(fig.~\ref{expansion}).  In an expansion fan the density and   
temperature decrease and the Mach number increases.  The expansion fan
is a smooth transition radiating from the bottom corner of the wedge.

The flow was computed numerically in a three-dimensional MD simulation
(fig.~\ref{wedge_shock}(b)) and in a two-dimensional finite difference 
simulation of
the inelastic continuum equations (fig.~\ref{wedge_shock}(c)). The two
simulations yield results for the horizontal component of velocity in  
qualitative accord with experiment: a shock forms at the tip of the
obstacle, and behind the shock the flow is compressed, has a higher  
temperature, and lower mean velocity.  Quantitative comparisons among
the methods are plotted for values of the fields along the dashed line
shown in fig.~\ref{wedge_shock}(a).

\begin{figure}
\begin{center}
\includegraphics[width=135mm]{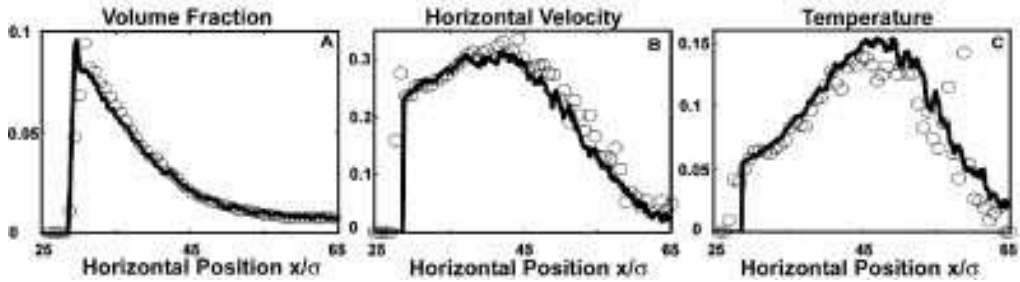}
\caption{Shock profiles for granular flow past a wedge measured in an
experiment (circles) are compared with results from molecular
dynamics (solid lines): (a) volume fraction, (b) horizontal component 
of the velocity, and (c) temperature.  The profiles are taken along
the dashed line in fig.~\ref{wedge_shock}.  (From
~\cite{rericha02})}\label{wp_friction}
\end{center}
\end{figure}

Three parameters were adjusted in the MD simulation to achieve the    
agreement with the experiment shown in figs.~\ref{wp_friction}.  The 
same coefficient of restitution $e_0=0.97$ and friction coefficient   
$\mu=0.15$ were used to model interparticle and particle-wall
collisions.  The initial conditions of the experiment were modeled by
placing particles into the top of the cell at a constant rate.
Incoming particles were placed randomly at the top of the cell with a
mean downward velocity measured from the experiment, and fluctuations
were chosen from a Gaussian distribution determined by the measured  
temperature. An additional parameter $\alpha$, defined as the ratio of
temperature perpendicular to the wall to that parallel to the wall,  
was set to 0.8.  These parameters, which were not measured in the  
experiment, were adjusted to provide agreement in the full flow
fields, including the free-stream velocity.

Results from the MD simulation are compared with experiment in
fig.~\ref{wp_friction} for the volume fraction, horizontal velocity
component, and temperature.  The agreement is quite good with a root 
mean square difference between experiment and simulation of less than
2\% for the volume fraction and velocity fields and 10\% for the 
temperature field.

The simple geometry and steady state behavior of the experiment
provided a good system for testing the inelastic continuum equations.
However, a full three-dimensional simulation of the experiment was   
found to be time prohibitive.  Instead, the equations were solved on a
two-dimensional grid; consequently the simulation could not capture the
interaction of particles with the confining glass side walls.
Frictional collisions with the side walls strongly affected the flow 
in the experiment and in the fully three-dimensional MD simulations.
The average downward acceleration of a single particle falling between
the two glass plates in the experiment was $8.9$ ${\rm m/s^2}$, while
the same particle falling outside the cell accelerated with the
expected $9.8$  $\rm m/s^2$.

The continuum equations were numerically solved by a second-order    
accurate, finite difference method.  The only fit parameter in the   
equations was the coefficient of restitution, which was set to the
same value of $e_0$(0.97) used in the MD simulation.  Boundary
conditions at the inlet were determined by the experiment and at the
outlet were free.  Slip velocity boundary conditions were used along 
the wedge boundary.  The heat flux at the wedge was taken to be      
proportional to the local $\nu$ and $T^{3/2}$~\cite{jenkins97}.  Euler
time stepping was used to increment the simulation until the flow
reached a steady state where the horizontally averaged mass flux was
constant to 0.01

\begin{figure}
\begin{center} \includegraphics[width=135mm]{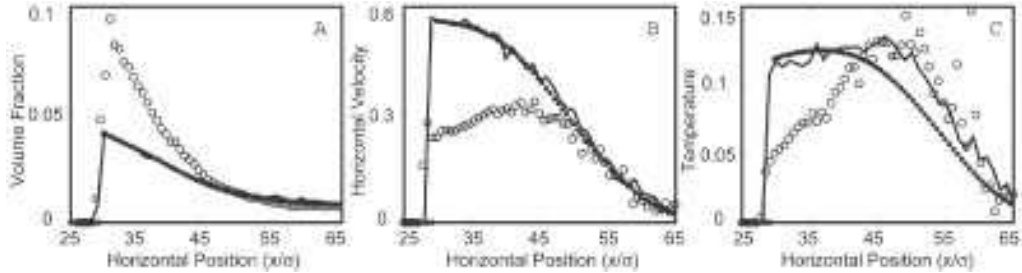}
\caption{Comparison of shock profiles for granular flow past a wedge
obtained from molecular dynamics (solid lines) and inelastic continuum
equations (dotted line), assuming no friction.  (a) Volume fraction,
(b) horizontal velocity profile, and (c) temperature along the dashed
line in fig.~\ref{wedge_shock}(a).  Experimental measurements
(open circles) show similar qualitative behavior but disagree
quantitatively.  The difference between the simulations and the
experiment is due to wall friction. (From ~\cite{rericha02})}
\label{wp_no_friction}
\end{center}
\end{figure}

Experiment and continuum simulation showed similar behavior, but
the shape of the curves differed and the magnitudes of the fields
disagreed by as much as a factor of two (fig.~\ref{wp_no_friction}).
This disagreement was attributed to the frictional drag of the confining side
walls in the experiment.  A three-dimensional simulation of the
inelastic continuum equations with viscous boundary conditions along
the side walls should agree better with the experiment
~\cite{goldshtein03}.

Molecular dynamics simulations were done with
wall drag neglected for comparison with the two-dimensional
simulation of the continuum equations.  The two simulations agreed
remarkably well in all regions of the flow except within 5$\sigma$ of the wedge 
tip.  Near the tip of the wedge, the two 
simulations disagreed due to different boundary conditions.  The agreement 
between the two simulations in the bulk of the flow 
confirms the applicability of the continuum description for granular flows. 
The disagreement with the experiment emphasizes once again the
importance of including friction in a continuum description, both in
the equations and in the boundary conditions.

\subsection{Shock waves in a vertically oscillating layer}

\begin{figure}
\begin{center}
\includegraphics[width=70mm]{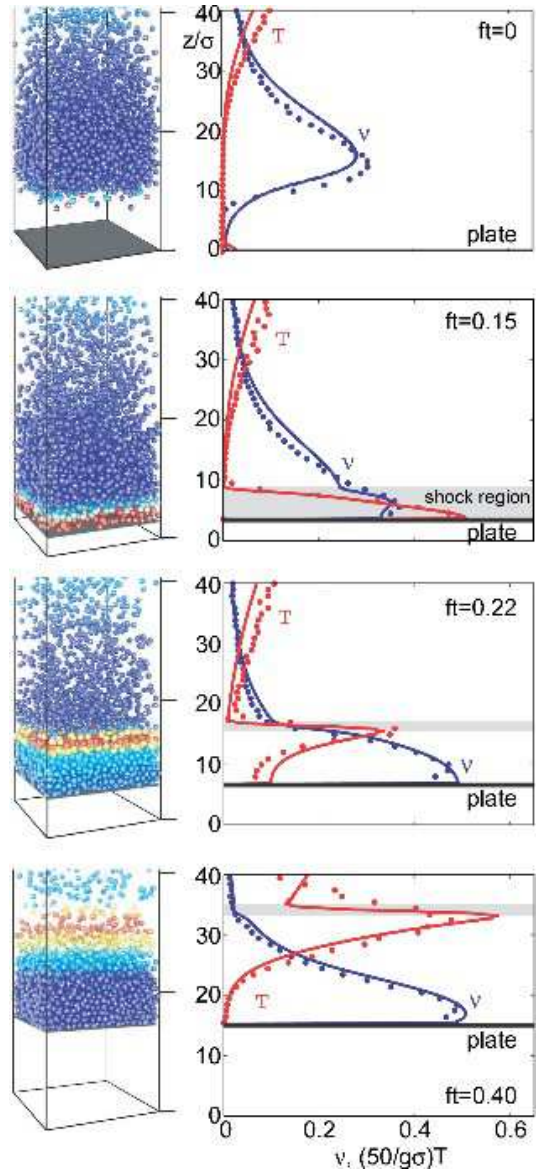}
\caption{Dimensionless temperature $T/g\sigma$ and volume fraction  
$\nu$ as functions of the dimensionless height $z/\sigma$ at four
times $ft$ in the oscillation cycle.  For each time, a snap shot from
the MD simulation is shown in the left column, with individual
particles color coded according to temperature: high T in red, low T 
in blue, and the bottom plate of the container shaded solid gray.  The
right column shows horizontally averaged $\nu$ (blue) and $T/g\sigma$ (red) for the
same four times.  The plate is shown as a horizontal black solid line,
results from MD simulation are shown as dots, and continuum results 
are solid lines (From ~\cite{bougie02}). }\label{prop_shock}
\end{center}
\end{figure}

Section 2 described the patterns that form when a layer of granular
materials is vertically oscillated.  Shocks play an important role in
this system~\cite{bougie02}.  Each time the layer collides with the
plate a shock forms and propagates through the layer, transmitting 
energy upward through the layer to the surface.  The
striking similarity of the granular patterns to their counterparts in 
continuum systems strongly suggests that a continuum 
description of
granular flows could prove useful. If this is true, the continuum    
description must also capture the shock dynamics in the bulk of the   
layer.

Here we compare inelastic, frictionless, fully three-dimensional MD and
continuum simulations.  In order to focus on the formation and
propagation of the shock wave, pattern formation is intentionally
suppressed by considering a container smaller than one wavelength of
the pattern in either horizontal direction.  We focus on one point in
the phase space shown in fig.~\ref{phasediagram}: $\Gamma=3$ and
$f^*=0.42$ for a layer with depth $h=9\sigma$.  For this set of
parameter values, a vibrated layer in a larger cell would have a
$f/2$ stripe pattern.

   In both simulations periodic boundary conditions were used in the two 
horizontal directions and impermeable boundary conditions, $u_z=0$, were 
applied 
at the 
plate.  The additional boundary conditions required for the continuum 
simulations were taken from the MD simulation.   In the MD simulation, the 
vertical derivatives at the plate were negligible throughout most of the 
cycle.  For simplicity, the continuum 
simulation assumed $\partial u_x/\partial z =0$,  $\partial u_y/\partial z 
=0$, and $\partial T/\partial z=0$ at all times in 
the cycle.

\begin{figure}
\begin{center} \includegraphics[width=60mm]{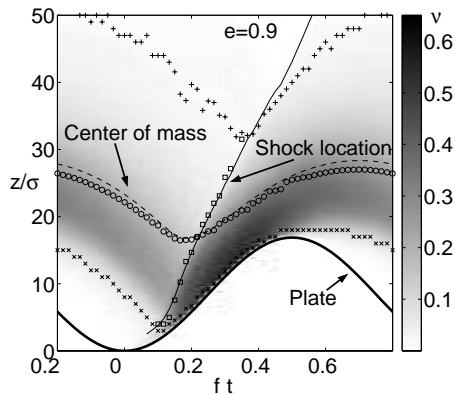}
\caption{Location of the shock (solid line for continuum, squares for
MD) and the center of mass of the layer (dashed line for continuum,
circles for MD) as a function of time $ft$ during one cycle of the
plate (thick solid line) for particles with $e_0=0.90$.  The plot is
shaded according to the volume fraction from the continuum simulation,  
so that high volume fraction is dark and low volume fraction is light.
The ``top'' and the ``bottom'' of the layer from MD (when the volume
fraction drops to less than 4\% of that for random close packed
particles) are shown as +'s and X's respectively.  The material below
the shock is compressed as compared to the region above the
shock, as can be seen from the shading. From(~\cite{bougie02}).}
\label{shock_location}
\end{center}
\end{figure}

The evolution of the shock wave throughout a plate cycle is shown in 
figs.~\ref{prop_shock} and ~\ref{shock_location}.  The dynamics of the 
cycle occurs in 
the time 
interval between $ft=0$ and one cycle later, $ft=1$.

At $ft=0$ the container is at its minimum height.  The layer, having  
been thrown off the plate in the previous cycle, now falls towards the
plate.  Inelasticity has dissipated most of the energy so that the
layer's temperature is nearly zero.  The Mach number of the layer with 
respect to the plate is much greater than one.  The MD and continuum 
simulation show similar behavior in the $v$ and $T$ fields.   

At $ft=15$ the layer begins to collide with the plate.  A shock wave
forms, separating the region near the plate where $\nu$ and $T$
increases from the undisturbed region still falling towards the plate.

At $ft=0.22$ the shock wave is moving through the layer.  The compressed
region continues to grow.  Collisions between particles in this high
density region cause the layer to cool behind the shock, creating a lower
temperature near the plate.

At $ft=0.40$ the shock has propagated through the layer and into the very
dilute region above the layer.  At this time, the plate is approaching its
maximum height and the layer begins to leave the plate as the downward
plate acceleration exceeds $g$.  The layer continues to cool behind the
shock, setting the stage for the next oscillation.

The MD and continuum simulations show good agreement throughout the cycle,  
despite the presence of large 
spatial gradients and a strong time dependence.  In the dilute regimes above 
and below the layer, numerical solutions of the inelastic continuum equations 
are unstable unless artificial dissipation is added~\cite{bougie02}, 
following 
the example from numerical solutions of Knudsen gases.~\cite{roache76}.  
The effect of the extra dissipation is most pronounced in the falling layer and 
accounts for the disagreement between MD and continuum at the top of the cell.  

Inelastic collisions between particles are the distinguishing
characteristics of a granular gas, but few studies have examined how
granular flow properties depend on the restitution coefficient.
Simulations for the oscillating layer were modified to study the   
effect of varying $e_0$ on the propagation of the shock.  The initial
conditions for this numerical experiment were taken for a layer with
$e_0=0.99$, using the same parameters as in the above discussion.  At   
$ft=0.33$, when the center of mass layer was near its maximum height  
above the plate, the coefficient of restitution was suddenly changed,
which changed the subsequent evolution of the shock.

As before, when the layer hits the plate it compresses and forms a
shock that propagates through the layer.  The smaller the value of
$e_0$, the faster the layer cools and compacts; for small $e_0$, the
layer remains very compact throughout the cycle and leaves the plate
almost as a solid body.  For higher values of $e_0$, the layer dilates
quickly after each collision with the plate.  The maximum height of
the center of mass in a cycle increases with increasing $e_0$.

\begin{figure}
\begin{center}
\includegraphics[width=60mm]{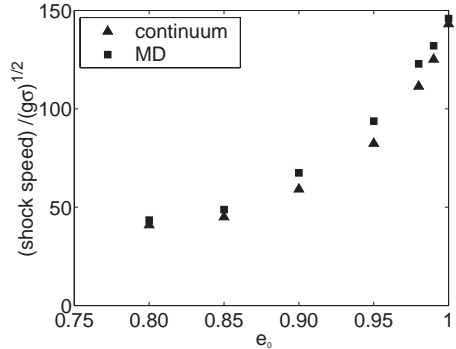}
\caption{Average dimensionless shock speed,
$v_{shock}/\sqrt{g\sigma}$, in the reference frame of
the plate.  $v_{shock}$ is calculated as the average
speed of the shock from when the shock is formed until
it leaves the layer. (From ~\cite{bougie02})}\label{shock_speed}
\end{center}
\end{figure}

The speed of the shock (eq.~\eqref{eq:speed}) depends on both the
temperature of the flow and on its density.  Since the density and 
temperature of the flow change throughout the cycle, so does the shock
speed as the shock propagates through the layer.  The behavior of the
average speed of the shock as a function of inelasticity is shown in    
fig.~\ref{shock_speed}.  For small values of $e_0$ the shock speed    
asymptotes to a fixed value of $47\sqrt{g\sigma}$.  The shock speed  
monotonically increases with increasing $e_0$.  The special case of
elastic particles appears to match with the limit of $e\rightarrow 1$,
suggesting that there is no qualitative difference in shock
propagation through elastic and inelastic gases.

For both problems we have considered -- granular flow past a wedge and
a vertically oscillating granular layer -- numerical solutions of the
inelastic continuum equations of Jenkins and Richman agree well with  
MD simulations for frictionless particles.  The continuum equations
were derived for a weakly dissipative, low density, frictionless
granular gas, assuming small gradients in the flow fields in both
space and time.  Nevertheless, the equations capture the evolution of
a shock through a dense, inelastic oscillating layer, and
qualitatively capture the properties of a shock formed in flow past a
wedge.  With more research on boundary conditions and the
incorporation of friction, these continuum equations show great
promise.

\section{Discussion}

More than one thousand papers have been published on granular
materials since de Gennes brought the subject to the attention of
physicists, and Bak's work (1987) on self-organized criticality
stimulated interest in sand piles~\cite{bak}.  However, much remains to 
be done to achieve a level of understanding of granular media
comparable to that for fluids and crystalline solids.  Experiments
and simulations have investigated a wide range of problems including  
the angle of repose~\cite{jaeger89} and internal structure of sand
piles~\cite{mehta94,vanel99}, shear forces in Couette-Taylor
flows~\cite{howell98,bocquet01}, convection due to temperature
gradients~\cite{hong95,ramirez00,ehrichs95} and due to
buoyancy~\cite{cordero03}, flows in a rotating
drum~\cite{shen02,hill03}, and chute
flows~\cite{sanders91,nott92,forterre01,campbell85a}.  Much
of the research has concerned granular media as a solid where particles 
are in continuous contact,  
while this chapter has concerned rapid granular flows (the
``collisional regime'') where inertial effects are important and force
chains do not play a major role.  We have further limited the
considerations to particles that interact only on contact.

Understanding flows of grains that interact only on contact would seem
at first to involve a straightforward application of Newton's
laws. However, the energy loss in collisions complicates the
application of standard statistical methods. We have focused on
two systems that hint at the rich variety of phenomena exhibited by 
granular media in the collisional regime: a vertically vibrated
granular layer and supersonic granular flow past an obstacle. These   
two problems were chosen because they are amenable to direct 
comparison of experiment,
molecular dynamics simulations, and continuum theory.

A vertically vibrated granular layer spontaneously forms spatially
extended patterns.  Experiments and MD simulations reveal that the
collective motion of grains arises due to dissipative collisions
between particles, and does not require mediation by an interstitial
gas or side walls. For square granular patterns, an approach   
intermediate between molecular dynamics and continuum models has been 
found to describe the dynamics of the lattice pattern: a collection of
particles that form a peak (an oscillon) is like an atom in a
crystalline lattice. The modes of the granular lattice obey the
dispersion relation for a two-dimensional lattice, and the granular
lattice even forms defects and melts in the same way as a
two-dimensional crystal of atoms (Section 2).

The spatial patterns formed by oscillating granular layers exhibit  
marked similarities to those observed in continuum nonequilibrium
systems such as convecting fluids and oscillating liquids (Section    
5). Further, the cross-roll and skew-varicose instabilities observed  
in thermal convection in a fluid and interpreted in terms of the
hydrodynamic equations (more specifically, the Boussinesq equations)
have also been observed in oscillated granular layers.  Various
amplitude equation models have been found to describe granular
patterns and their instabilities.  Even the subtle effects of noise on
the transition from conduction to convection in fluids have been found
also in oscillating granular layers near the onset of the transition
from a flat layer to a square pattern.

The striking similarities of granular patterns to those found in
nonequilibrium continuum systems and in experiments on granular flows
under shear and in rotating drums suggest that granular gases may be
describable by continuum theory.  Inspired by these observations,
researchers have proposed many continuum descriptions.  The   
descriptions differ in the particle properties included in the
collision model; for instance, collision models can be 
frictionless~\cite{lun84,jenkins85} or can account for friction
between particles~\cite{lun91,luding98}.  Equations of motion obtained by 
Goldshtein and Shapiro include, in
addition to the terms in the Navier-Stokes equation, a
term accounting for heat transport by density
gradients~\cite{goldshtein95}.  A Model presented by Bocquet {\it et 
al.}~\cite{bocquet01} includes corrections to the viscosity due
to velocity correlations.  None of these models has been definitively
established.

We have compared predictions of continuum equations derived by
Jenkins and Richman with experiments on shocks in vibrating layers and flow 
past an obstacle.  For both geometries, numerical solutions of the
inelastic continuum equations agree well with results from MD
simulations of smooth (frictionless) inelastic spheres.  However,  
comparisons of the continuum equations with experiment and with MD
simulations for particles with friction have demonstrated the crucial role
of friction in granular flows.  For continuum equations to achieve
quantitative predictive power, the effects of friction between
particles and between particles and boundaries must be included.

Derivations of granular hydrodynamic equations have thus far assumed
weak dissipation and small particle volume fractions.  Future work
should extend theory to higher densities and larger dissipation.
Because inelastic collisions dissipate temperature, granular flows 
frequently coexist with solid-like sand piles.  A major challenge is
the development of theory that bridges the gap from the collisional regime 
to 
the quasi-static regime where particles are always
in contact.

As continuum equations of motion become better established, it will become
possible to exploit the power of the continuum description.  Continuum
models for larger and denser systems may reveal new
phenomena.  Continuum simulations are better suited to time-dependent
flows than MD simulations.    Linear
and nonlinear stability analyses could provide insight into
bifurcations, just as a century of stability analyses of the Navier-Stokes
equations has given insights into diverse fluid flow phenomena.
Stability analyses have been conducted for simplified continuum models
(see, e.g.,~\cite{bizon99}), but thus far no stability analysis has
been conducted for a realistic set of granular hydrodynamic equations.
Further, given a set of granular equations like the Navier-Stokes equations,
it should be possible to derive amplitude equations, which can
yield a better understanding of instabilities in granular flows.

The experiments and theory presented in this chapter involve spheres
of uniform size, while industrial applications usually involve a wide
range of particle sizes and shapes.  Experiments and simulations on
flows with two particle sizes show an additional richness to granular 
flow phenomena such as size
segregation~\cite{moon04,hill01,williams76,rosato87,metcalfe96}, 
nonequipartition of energy~\cite{feitosa02,alam03}, and increased normal
stress~\cite{alam03}. Much theoretical and experimental work is needed
on systems of particles with a range of sizes and shapes.

Air friction is usually neglected in simulations and in the
interpretation of experiments, whether or not the experiments are conducted in
vacuum. In contrast, in granular systems in industry, air friction and
buoyancy are often important, although the air effects can be negligible
for large particles (say greater than 1 mm).  
The interaction of the interstitial fluid and sand leads to the 
development of sand dunes~\cite{bagnold} and sand 
ripples~\cite{duran00,vittori90} and the formation of heaps in 
vibrated layers~\cite{pak95}.   The 
inclusion of
interstitial flow in continuum theories for granular materials is another
challenge for future research on granular materials.

In conclusion, much remains to be done to establish a fundamental 
understanding granular flows. Future studies should seek more examples of 
granular flows
amenable to experiment, molecular dynamics simulations, and
continuum theory.  A concerted attack from three approaches should lead to  
a better understanding of outstanding issues concerning boundary conditions, 
the
incorporation of particle friction and velocity correlations into theory, and 
the
extension of theory to higher particle volume fractions and higher dissipation.

\section{Acknowledgments}

In the past two decades granular media have been studied by many
scientists.  For convenience, our examples have been taken largely
from research in the Center for Nonlinear Dynamics at The University of 
Texas at Austin (see
http://chaos.utexas.edu). We thank particularly Chris Bizon, Jonathan
Bougie, Francisco Melo, Daniel Goldman, W. D. McCormick, Sung Joon
Moon, Mark Shattuck, Jack Swift, and Paul Umbanhowar.  This research
was supported by the Engineering Research Program of the Office of
Basic Energy Sciences of the U.S. Department of Energy and the Texas
Advanced Research Program.

\newcommand{\bibnamefont}{\MakeUppercase}

\end{document}